\documentclass[aps,prx,twocolumn,superscriptaddress,nofootinbib]{revtex4-2}

\usepackage{graphicx}
\usepackage{dcolumn}
\usepackage{bm}
\usepackage{amsmath,amssymb}
\usepackage{xcolor}
\usepackage{physics}
\usepackage{booktabs}
\usepackage{comment}

\usepackage{mfirstuc}
\usepackage[normalem]{ulem}

\usepackage{hyperref}
\hypersetup{
  colorlinks=true,
  linkcolor=blue,
  citecolor=blue,
  urlcolor=blue
}

\newcommand{\addComment}[2]{
  \expandafter\newcommand\csname #1\endcsname[1]{{\bf \color{#2} \capitalisewords{#1}:\,##1}}
  \expandafter\newcommand\csname #1cor\endcsname[2]{{\color{#2} \capitalisewords{#1}:\,\st{##1}{\bf ##2}}}
  \expandafter\newcommand\csname #1color\endcsname{#2}
}
\addComment{cris}{blue} 

\begin{document}


\title{Physics-Aware, Shannon-Optimal Compression via Arithmetic Coding \\ for Distributional Fidelity}

\author{Cristiano Fanelli}
\email{cfanelli@wm.edu}
\affiliation{School of Computing, Data Sciences, and Physics, William \& Mary, Williamsburg, VA, USA}

\date{\today}


\begin{abstract}
Assessing whether two datasets are \emph{distributionally consistent} is central to modern scientific analysis, particularly as generative artificial intelligence produces synthetic data whose fidelity must be validated against real observations in increasingly high-dimensional settings. Existing approaches are typically relative: they determine whether one dataset is more consistent with a reference than another, but do not provide a physically grounded absolute standard for fidelity.
We propose an information-theoretic approach in which lossless compression via \emph{arithmetic coding} provides an operational measure of dataset fidelity under a physics-informed probabilistic representation. Datasets sharing the same underlying physical correlations admit comparable optimal descriptions, while discrepancies-arising from miscalibration, mismodeling, or bias-manifest as an irreducible excess in codelength relative to the Shannon-optimal limit defined by the physics itself.
This excess codelength defines an \emph{absolute} fidelity metric, quantified directly in bits. Unlike conventional measures, which lack an intrinsic scale, zero excess provides a well-defined and physically meaningful target corresponding to consistency with the underlying distribution. We show that this metric is global, interpretable, additive across components, and asymptotically optimal, with differences in codelength corresponding to differences in expected negative log-likelihood under a common reference model. As a byproduct, our approach achieves improved compression relative to standard general-purpose algorithms such as \texttt{gzip}.
These results establish arithmetic coding not merely as a compression tool, but as a measurement instrument for absolute, physics-grounded assessment of distributional fidelity.
\end{abstract}

\maketitle


\section{Introduction}
\label{sec:introduction}

Assessing whether two datasets are described by the same underlying probability distribution is a foundational problem across many disciplines \cite{lehmann2005testing, wasserman2004all, cover1999elements, gretton2012kernel, rachev2013methods}. In the context of generative artificial intelligence, this question arises naturally when evaluating how faithfully synthetic data produced by generative models reproduce the distribution of the original data \cite{goodfellow2014generative,theis2015note, kynkaanniemi2019improved,naeem2020reliable}. In the physical sciences, closely related challenges appear in comparisons between Monte Carlo simulations and data collected under varying experimental conditions, as well as in detector calibration, data validation, and the assessment of the fidelity of approximate or generative simulation techniques \cite{kansal2023evaluating,cranmer2020frontier, paganini2018calogan, butter2021ganplifying, karniadakis2021physics, deistler2025simulation}.
As modern experiments generate increasingly large, high-dimensional, and multimodal datasets, traditional approaches to distributional comparison face growing limitations. Methods based on handcrafted test statistics, low-dimensional summaries, or explicitly parameterized likelihoods often rely on design choices that become difficult to justify or interpret as dimensionality and complexity increase, and whose behavior may be dominated by modeling assumptions rather than by intrinsic properties of the data.

A wide range of statistical tools have been developed to address distributional consistency, including divergence-based measures, kernel-based distances, embedding-space metrics, and classical goodness-of-fit tests. While powerful in specific settings, these approaches generally require the explicit choice of a test statistic, kernel, feature space, or approximate likelihood, introducing assumptions that are external to the data representation itself. In high-dimensional and multimodal regimes, such choices can strongly influence sensitivity and interpretation \cite{arjovsky2017towards}.
Divergence-based measures, such as the Kullback–Leibler and Jensen–Shannon divergences, provide principled information-theoretic notions of distributional discrepancy and admit well-defined axiomatic and geometric interpretations \cite{cover1999elements,lin2002divergence,fuglede2004jensen}. However, these divergences are not directly observable from finite samples and typically require explicit density models, parametric assumptions, or variational approximations to be estimated in practice \cite{arjovsky2017towards,nowozin2016f}.

Sample-based distances, including kernel methods such as the Maximum Mean Discrepancy or optimal-transport–based metrics, avoid explicit density estimation but depend on externally specified structures—such as kernel bandwidths or cost functions—that are not uniquely determined by the physical representation of the data \cite{gretton2012kernel,ramdas2015decreasing}. Embedding-based fidelity metrics assess consistency in a learned feature space rather than at the level of the original data distribution \cite{gretton2012kernel,arjovsky2017towards}, while classical goodness-of-fit tests—including binned and unbinned likelihood-based tests—rely on selected observables or binning schemes and, in the large-sample limit, may reject the null hypothesis for arbitrarily small perturbations, quantifying statistical detectability rather than practical distributional fidelity \cite{lehmann2005testing,wasserman2006all,bickel2015mathematical,van2000asymptotic}.

In this work, we use \emph{lossless compression} as an operational probe of distributional fidelity. Rather than introducing an auxiliary distance, classifier, or feature space, we interpret the \emph{excess codelength} between datasets as a representation-conditional statistic that directly quantifies distributional mismatch. For a fixed, physics-informed probabilistic representation, lossless compression provides a canonical and intrinsic realization of distributional comparison.

Arithmetic coding (AC)~\cite{witten1987arithmetic, mackay2003information} plays a central role in this
framework. As a lossless compression technique, AC provides a constructive realization of
Shannon-optimal encoding for a given probability assignment. We employ a \emph{physics-aware}
arithmetic codec whose probabilistic structure reflects known features of the physical processes
underlying the detector response. Given a probabilistic model $q(x)$, arithmetic coding produces a
binary description whose length approaches $-\log_2 q(x)$, up to finite-precision effects,
establishing a direct and architecture-independent mapping between probability and achievable
codelength. In this work, AC is used not as a production codec, but as a \emph{reference instrument}
that renders probabilistic structure observable through measured codelengths.

The present approach is related in spirit to the Minimum Description Length (MDL) principle~\cite{rissanen2009model}, in which compression is used for model selection by balancing goodness of fit against model complexity. In this work, by contrast, the probabilistic representation is fixed based on physical considerations, and compression is used diagnostically to quantify distributional mismatch between datasets rather than to select among competing models.
Recent work on learned and neural compression focuses on optimizing representations to achieve higher compression rates \cite{balle2018variational,townsend2019hilloc}, while in contrast we hold the probabilistic representation fixed and use lossless compression to expose missing or broken physical correlations through irreducible excess codelength.

Within our framework, calibration and consistency tests are formulated directly in terms of achieved codelength. Data that respect the correlations encoded in the reference representation compress efficiently, while violations of those correlations incur an expected excess codelength. 
The resulting diagnostic is global—sensitive to the full joint distribution over the chosen representation rather than to selected observables or low-dimensional projections—interpretable, with deviations expressed directly in bits, and additive, allowing in this work contributions from detector subsystems or data components to be accumulated and compared consistently.

From a statistical perspective, increasing sample size improves the stability of this diagnostic
by reducing finite-sample fluctuations and driving the measured average codelength toward its
asymptotic value,
\begin{equation}
\lim_{N \to \infty} \frac{1}{N} \sum_{i=1}^{N} \ell(x_i)
= H(p) + D_{\mathrm{KL}}(p \| q),
\end{equation}
where $p$ denotes the true data-generating distribution and $q$ the reference probabilistic
representation. From a computational standpoint, arithmetic coding scales linearly with the number
of encoded symbols, and practical implementations may employ blocking, streaming, or parallelization without altering the underlying information-theoretic interpretation.

Lossless compression does not replace existing fidelity diagnostics, but provides a
representation-consistent and information-theoretic realization of them. Unlike kernel-based
distances or embedding-space metrics, compression does not rely on externally defined feature
spaces or test functions: all correlations present in the data contribute automatically to the
achievable codelength. In this sense, compression probes fidelity at the level of the full data
distribution.
Using controlled perturbations and independent reference samples, we demonstrate that
physics-aware arithmetic coding provides a calibrated and interpretable diagnostic of distributional
inconsistency for multimodal detector readout data, complementary to conventional distributional
tests. Compression is thus elevated from a data-reduction technique to a quantitative instrument
for validating physical structure in complex datasets.

The structure of this paper is as follows. Section~\ref{sec:data} describes the dataset used in this work. Section~\ref{sec:framework} introduces the conceptual framework. Section~\ref{sec:analysis} presents the results and their interpretation. Section~\ref{sec:conclusions} summarizes the main conclusions.

\section{Data and Representation} 
\label{sec:data}

We use simulated electromagnetic calorimeter data from the CLAS12 detector, including the PCAL, ECIN, and ECOUT subsystems, as described in Refs.~\cite{ungaro2020clas12, ungaro2024geant4}.
The calorimeter consists of alternating lead and scintillator layers read out in three stereo views (U/V/W) rotated by approximately $60^\circ$, providing complementary projections of the transverse shower profile.

Each event is represented by integer-valued detector readout and particle-level quantities:
\texttt{hits\_adc} and \texttt{hits\_strips} store arrays of shape $(N,9,20)$ with up to 20 hit slots per layer; 
\texttt{max\_adc} and \texttt{max\_strips} have shape $(N,9)$ and record the maximum response per layer;\footnote{\texttt{max\_adc} and \texttt{max\_strips} can be derived from \texttt{hits\_adc} and \texttt{hits\_strips} and for this reason are not used in this work.} 
and \texttt{part\_input} has shape $(N,3)$ and contains the particle momentum components $(p_x,p_y,p_z)$.
The nine layers correspond to PCAL (U/V/W), ECIN (U/V/W), and ECOUT (U/V/W).
In this study, we consider data from a single calorimeter sector; the nine readout layers have different numbers of physical strips, namely 68 (PCAL-U), 62 (PCAL-V), 62 (PCAL-W), and 36 strips for each of the ECIN and ECOUT U/V/W layers.
For these studies, we use a dataset comprising $\mathcal{O}(10^6)$ events.
Padding is encoded using \texttt{strip} $= -999$, which is paired (by construction) with the unphysical value \texttt{adc} $= -999$; this invariant is verified and enforced throughout.
Hit-slot indices carry no geometric meaning across layers or views: correlations arise solely from the joint statistics of the readout, not from positional alignment of slots. 

The dataset is inherently multimodal, combining sparse, heavy-tailed discrete detector responses, categorical strip identifiers with geometry-dependent correlations, layer-dependent occupancy patterns, and continuous particle kinematics:
\begin{equation}
\label{eq:dataformat}
    X = \{\texttt{hits\_adc}, \texttt{hits\_strips}, \texttt{part\_input}\},  
\end{equation}
No lossy preprocessing is applied to the stored dataset. All quantities are used at full integer precision, with no clipping, binning, or thresholding beyond the detector’s intrinsic
digitization. The representation is therefore lossless with respect to the recorded detector
readout and particle kinematics; any compression gain arises exclusively from exploiting statistical structure rather than from information removal.
This representation motivates the physics-aware factorization used by the codec and underpins the fidelity studies presented in Sec.~\ref{sec:analysis}.

\paragraph*{Dataset splits}

To disentangle model training, reference evaluation, and fidelity testing, we repeat independent random splits (70/30\%) of the same reference dataset $R$. The first split $(A^{(1)},B^{(1)})$ is used to prove invertibility of the test dataset in Sec. \ref{sec:invertibility} and to study compression properties in Sec. \ref{subsec:compression_ratio}; this split is also used to define the perturbed simulation in Sec. \ref{sec:fidelity}: a perturbed sample $C$ is constructed by applying a controlled transformation (\textit{e.g.},\ an ADC scale distortion) to $B^{(1)}$, ensuring that $C$ and $B^{(1)}$ share identical event indices while differing only through the imposed effect. A second independent split $(A^{(2)},B^{(2})$ provides an unbiased reference sample for statistical comparisons, avoiding reuse of the data that seeded the perturbation. Finally, a third split $(A^{(3)},B^{(3)})$ is used to create the training dataset $A^{(3)}$ to train the arithmetic codec, thereby fixing a reference probability model $q_{A^{(3)}}(x)$ that is in good approximation statistically independent of both $C$ and $B^{(2)}$. All fidelity tests are then performed by encoding $C$ and $B^{(2)}$ under the same codec trained on $A^{(3)}$. This separation ensures that observed codelength differences reflect genuine distributional inconsistencies rather than artifacts of training–testing overlap or sample reuse.
For notational simplicity, the dataset label \(^{(i)}\) is suppressed when it does not affect the interpretation.

\section{Conceptual Framework}
\label{sec:framework}

We now formalize the information-theoretic framework underlying our approach.
Our central premise is that \emph{compression itself is not the source of physical understanding}: arithmetic coding is an optimal executor of a given probability model, while physical structure enters only through the probabilistic representation used to assign probabilities to the data.

The achieved codelength therefore provides an operational measurement of how well a fixed, physics-informed model explains a dataset, expressed directly in bits.

\subsection{Arithmetic Coding and Shannon Optimality}
\label{sec:framework:ac}

Arithmetic coding~\cite{witten1987arithmetic,mackay2003information} is a lossless entropy coding algorithm that maps a sequence of discrete symbols to a binary string whose length approaches the negative log-probability of that sequence under a specified model.
Let $x=(x_1,\dots,x_T)$ be a sequence with model probability
$q(x)=\prod_{t=1}^{T} q(x_t\mid x_{<t})$.
Ignoring finite-precision termination effects, AC produces a code of length
\begin{equation}
\ell_q(x)\;\simeq\; -\log_2 q(x)
\;=\;\sum_{t=1}^{T} \big[-\log_2 q(x_t\mid x_{<t})\big].
\label{eq:ac_nll}
\end{equation}
In practice, AC is implemented with integer arithmetic and cumulative distribution functions (CDFs), which introduces a small, implementation-dependent overhead.
For a correct range codec, the achieved length satisfies
\begin{equation}
-\log_2 q(x)\;\le\;\ell_q(x)\;<\;-\log_2 q(x)+\varepsilon,
\label{eq:ac_overhead_bound}
\end{equation}
with $\varepsilon$ typically negligible at the scale of the results reported here.

\paragraph{Code length, entropy, and model mismatch.}
Consider data drawn from an unknown distribution $p(x)$ that are encoded using a (possibly mismatched) model $q(x)$.
Taking expectations under $p$, the average achieved length converges to the
\emph{cross-entropy} between $p$ and $q$,
\begin{equation}
\begin{aligned}
\mathbb{E}_{x\sim p}\!\big[\ell_q(x)\big]
&\;\simeq\;
\mathbb{E}_{x\sim p}\!\big[-\log_2 q(x)\big]
\;\equiv\;
H(p,q)
\\[0.3em]
&=\;
H(p) + D_{\mathrm{KL}}(p\Vert q)\, ,
\end{aligned}
\label{eq:cross_entropy_decomp}
\end{equation}
up to the small overhead in Eq.~\eqref{eq:ac_overhead_bound}.
Equation~\eqref{eq:cross_entropy_decomp} identifies the \emph{excess bits} beyond $H(p)$ with the model mismatch $D_{\mathrm{KL}}(p\Vert q)$.
A richer probabilistic structure can capture additional correlations present in the data, thereby reducing the intrinsic entropy of the modeled process and sharpening sensitivity to structured deviations.
However, at finite statistics this need not imply a shorter achieved codelength: increased model complexity, estimation error, or factorization overhead can increase the cross-entropy even as the underlying entropy decreases.
Any change in codelength therefore arises from the probabilistic representation of the data distribution-not from modifications of the coding algorithm itself.

\subsection{Physics-aware Probabilistic Representations}
\label{sec:framework:factorizations}

Our codec operates on a fixed, discrete representation of each event composed of (i) calorimeter hit data and (ii) an auxiliary particle-kinematics payload.
The probabilistic model used by arithmetic coding is specified through a set of discrete conditional distributions represented as CDF tables.
These CDFs can be obtained from empirical frequency estimates or from learned models; in both cases, they provide the probability assignment $q(\cdot)$ required by AC.

\paragraph{Unconditional factorization.}
For detector hits, we exploit the structured, sparse nature of the readout and use a physics-informed factorization that separates occupancy from the discrete hit attributes.
Denoting by $\mathrm{occ}$ the occupancy indicator for each layer--slot, and by $(\mathrm{strip},\mathrm{adc})$ the corresponding discrete values when occupied, we write
\begin{equation}
q(\text{hits})
\;\approx\;
\prod_{\text{layers},\,\text{slots}}
q(\mathrm{occ})\,
q(\mathrm{strip}\mid \mathrm{occ})\,
q(\mathrm{adc}\mid \mathrm{occ}),
\label{eq:factor_hits_uncond}
\end{equation}
which induces a sequential symbol stream for AC while preserving the causal structure implicit in the readout (occupancy first, then attributes).
Here, ``slots'' denote fixed index positions in the per-layer hit list used by the representation (up to a fixed maximum per layer), rather than physical detector channels.

\paragraph{Conditional factorization.}
To incorporate known causal dependence of the detector response on the particle kinematics, we also consider a conditional variant in which the hit distributions are conditioned on the particle momentum magnitude $|p|$ through a finite set of bins:
\begin{align}
q(\text{hits}\mid |p|)
&\approx
\prod_{\text{layers},\,\text{slots}}
q(\mathrm{occ}\mid |p|)\,
q(\mathrm{strip}\mid \mathrm{occ},|p|)\,
\nonumber\\
&\hspace{2.5em}\times
q(\mathrm{adc}\mid \mathrm{occ},|p|).
\label{eq:factor_hits_cond}
\end{align}
Conditioning is expected to reduce the intrinsic randomness of the hit stream \emph{as represented by the model} by capturing correlations between event kinematics and detector response.\footnote{Hit distributions are conditioned on $|p|$ by discretizing $|p|$ into a finite set of contiguous bins spanning the range $0$--$10~\mathrm{GeV}$, with bin edges chosen to balance kinematic coverage and statistical stability across the momentum range.} 
At the information-theoretic level, this replaces the marginal entropy $H(\text{hits})$ with the conditional entropy $H(\text{hits}\mid |p|)$ as the relevant entropy bound, even though the achieved cross-entropy---and thus the observed codelength---may increase at finite statistics.

\paragraph{Auxiliary kinematics payload.}
To isolate calibration sensitivity to detector-response mismodeling, we intentionally encode the kinematics using a generic, non-physical byte-level factorization.
Let $\vec p=(p_x,p_y,p_z)$ be stored as a fixed-width byte sequence $(b_1,\dots,b_M)$, where $M$ is the total number of bytes used to encode
the kinematic vector. We model
\begin{equation}
q(\vec p)\;\approx\;\prod_{m=1}^{M} q(b_m),
\label{eq:factor_part_bytes}
\end{equation}
treating kinematics as auxiliary information without exploiting physical correlations.
This design choice ensures that improvements in sensitivity arise from the hit model (unconditional vs.\ conditional), rather than from a more expressive kinematics codec.

\paragraph{Lossless compression and closure.}
Compression is lossless by construction: decoding inverts encoding exactly, returning identical integer arrays for all streams.
We validate this property through a closure test (encode $\rightarrow$ decode $\rightarrow$ bitwise equality), and separately quantify the finite-precision overhead via the difference between the achieved average codelength and the cross-entropy implied by the same CDF tables.

It should be noted that, within each layer–view and kinematic context, the ADC distribution is modeled conditional on hit occupancy but independent of the specific strip index. This factorization reflects the observation that the dominant correlations between energy deposition and detector geometry are mediated by event-level kinematics, while further conditioning on strip index leads to statistically fragile estimates with limited additional compression or sensitivity gains. 
The resulting model balances probabilistic expressivity, stability, and invertibility, providing a robust reference distribution for perturbation-based fidelity tests.

\subsection{Lossless Compression as a Likelihood-Scored Fidelity Diagnostic}
\label{sec:framework:lrtest}

Arithmetic coding provides an operational realization of negative log-likelihood under a \emph{fixed} reference model.
Let $q_{A}$ denote the probability model (CDF tables) learned from a reference dataset $A$ and then held fixed.
For a dataset $D=\{x_i\}_{i=1}^{N}$ encoded with this codec, the total codelength satisfies
\begin{equation}
L^{tot}_{A}(D) 
\;\equiv\;
\sum_{i=1}^{N}\ell_{q_A}(x_i)
\;\simeq\;
\sum_{i=1}^{N}\left[-\log_2 q_A(x_i)\right],
\label{eq:total_codelength}
\end{equation}
so that the average codelength estimates the cross-entropy between the empirical distribution of $D$ and the reference model:
\begin{equation}
\frac{1}{N}L^{tot}_{A}(D)
\;\approx\;
H(\hat p_D, q_A)
\;\equiv\;
\mathbb{E}_{x\sim \hat p_D}\!\left[-\log_2 q_A(x)\right].
\label{eq:cross_entropy_empirical}
\end{equation}
Here $\hat p_D$ denotes the empirical distribution associated with dataset $D$, while the unknown population distribution is denoted by $p$.
Given an independent baseline sample $B$ from the same population as $A$ and a perturbed (or synthetic) sample $C_\varepsilon$, we define the \emph{excess codelength}
\begin{equation}
\Delta L_{A}(\varepsilon)
=
 L_{A}(C_\varepsilon)
-
 L_{A}(B),
\label{eq:deltaL_def}
\end{equation}  
where $ L_{A}(S) \equiv L^{tot}_{A}(S)/|S|$ denotes the average codelength per event for a sample $S$ encoded with the fixed reference model $q_A$.
This quantity compares the \emph{mean} negative log-likelihood under the \emph{same} fixed scoring rule.
Crucially, Eq.~\eqref{eq:deltaL_def} does not require event-by-event pairing between $B$ and $C_\varepsilon$; it is a comparison of sample means under a shared reference model.

Interpreting $-\log_2 q_A(x)$ as a per-event codelength (in bits),
$\Delta L(\varepsilon) > 0$ indicates that events in the perturbed sample $C_\varepsilon$
are, on average, less typical under the fixed reference model $q_A$ than events in the
independent baseline sample $B$.
Equivalently,
\begin{align}
\Delta L(\varepsilon)
&\approx
H(\hat p_{C_\varepsilon}, q_A) - H(\hat p_B, q_A)
\nonumber\\
&=
\mathbb{E}_{x\sim \hat p_{C_\varepsilon}}\!\big[-\log_2 q_A(x)\big]
-
\mathbb{E}_{x\sim \hat p_B}\!\big[-\log_2 q_A(x)\big].
\end{align}
Here $\hat p_{C_\varepsilon}$ and $\hat p_B$ denote the empirical distributions of
$C_\varepsilon$ and $B$, respectively.
Thus, $\Delta L(\varepsilon)$ is the difference between two cross-entropies evaluated under
the same fixed reference model $q_A$—trained once on dataset $A$—with expectations taken
over independent samples from the baseline and perturbed datasets.
This yields a model-conditional consistency test that assesses whether $C_\varepsilon$
remains typical under the physical assumptions encoded by the reference codec,
rather than testing equality in an externally specified feature or embedding space.

In the fidelity studies (Sec.~\ref{sec:analysis}), statistical significance is assessed by
estimating $\Delta L(\varepsilon)$ on $K$ approximately independent event blocks and
applying a one-sided hypothesis test to the resulting blockwise estimates.
The outcome is expressed in bits per event, enabling a direct and additive interpretation
of the degree to which distributional structure is violated under the fixed reference
representation.

\section{Analysis and Results}
\label{sec:analysis}
\subsection{Invertibility}
\label{sec:invertibility}

\begin{figure*}[!]
    \centering
    \hspace{-0.9cm}
    \includegraphics[width=0.43\textwidth]{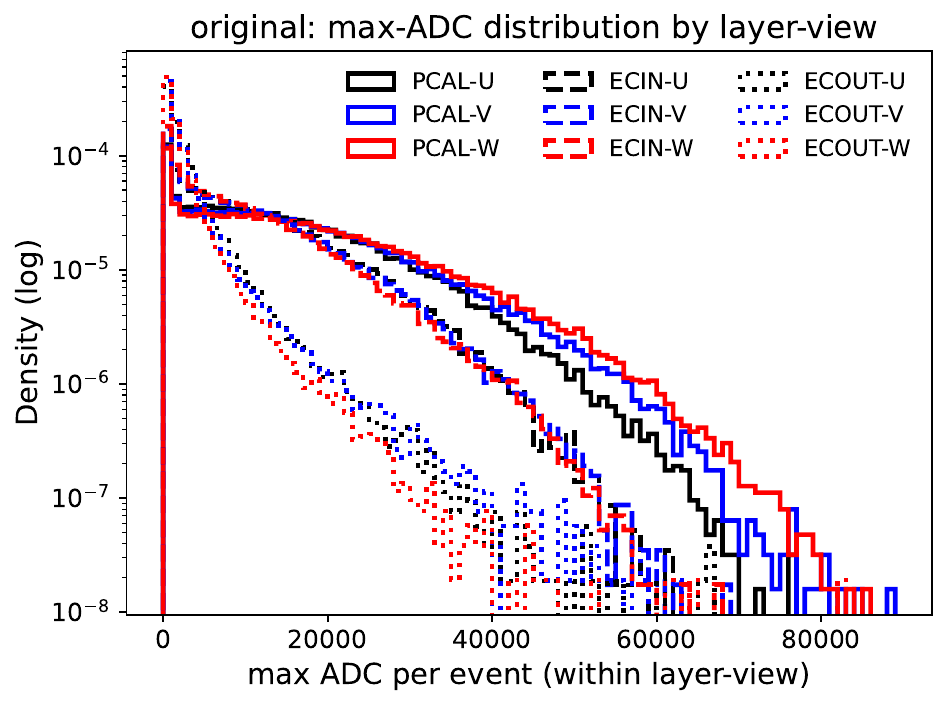}
    \hspace{0.65cm}
    \includegraphics[width=0.43\textwidth]{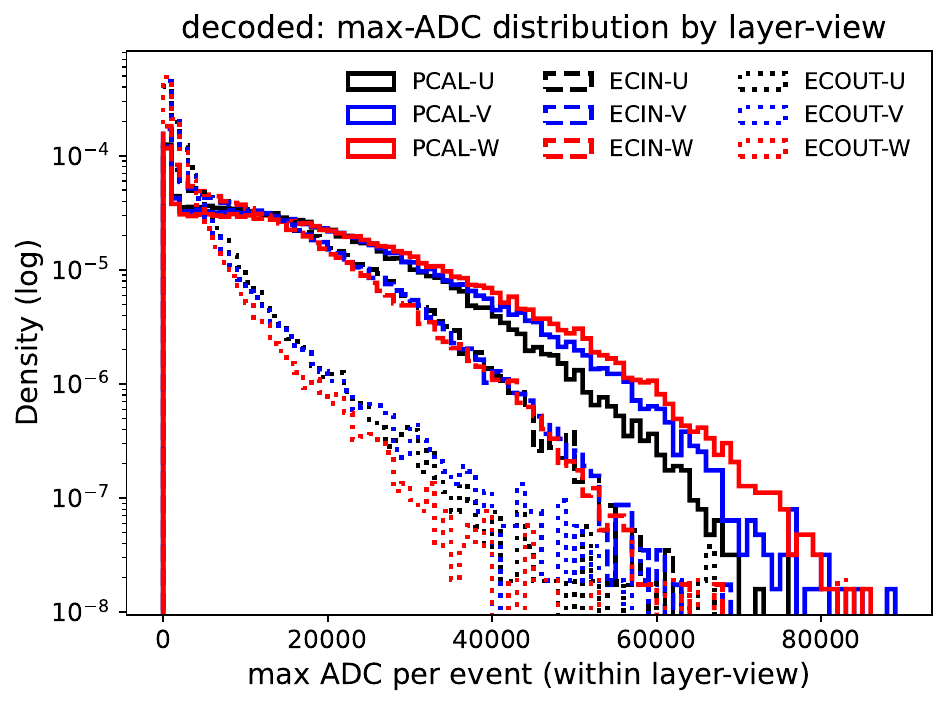}

    \vspace{0.3cm}

    \includegraphics[width=0.475\textwidth]{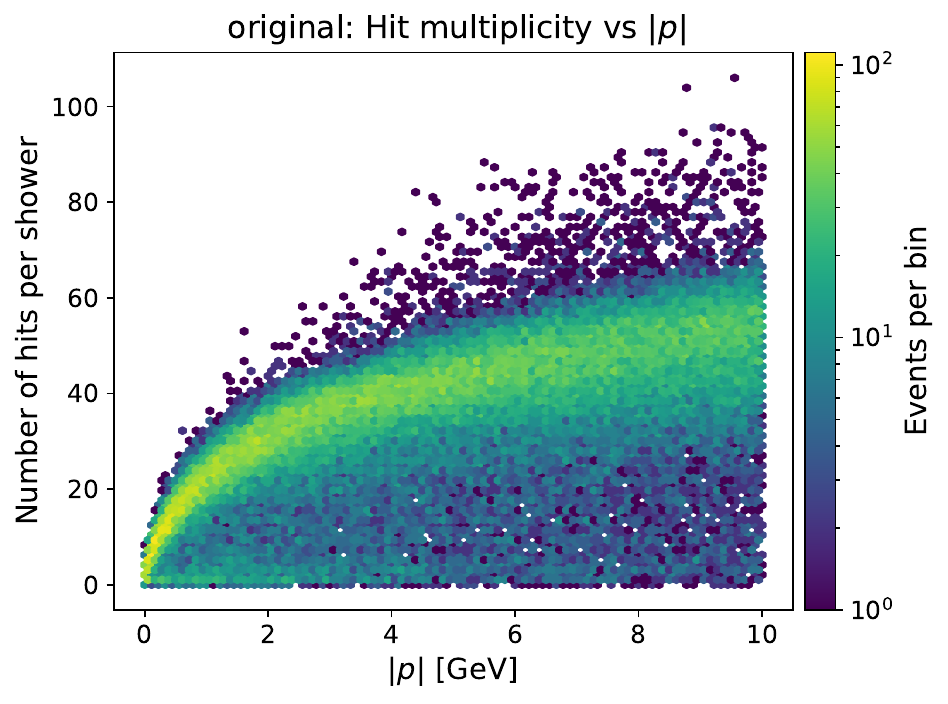}
    \includegraphics[width=0.475\textwidth]{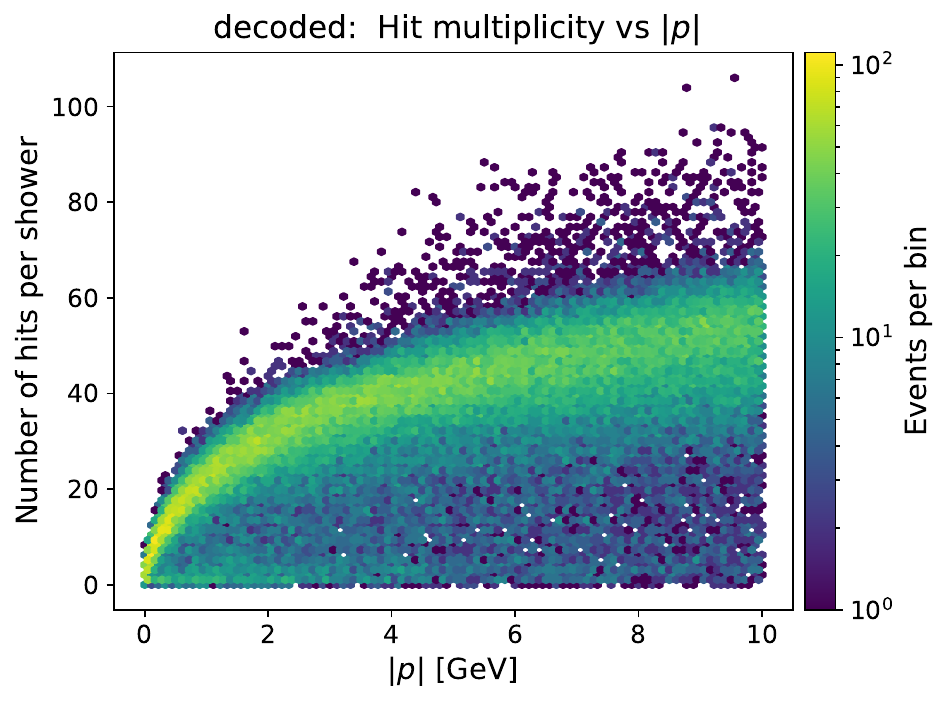}

    \caption{
    \textbf{Comparison between original and decoded detector-level observables.}
    (Top row) ADC distributions by detector layer for the original data (left) and the decoded data after compression (right).
    (Bottom row) Hit multiplicity as a function of particle momentum $p$ for the original (left) and decoded data (right). Arithmetic coding is lossless and invertible, and the exact agreement across all panels demonstrates that the compression–decompression cycle preserves both low-level and derived detector features. This result holds for both codecs (unconditional and conditional to the particle kinematics).
    }
    \label{fig:original_vs_decoded_observables}
\end{figure*}

To verify the lossless and invertible nature of the proposed arithmetic coding framework, we perform a direct comparison between detector-level observables computed from the original data and from the data obtained after a full compression--decompression cycle.
The dataset $R$ is split into disjoint training ($A$) and validation ($B$) subsets, with models trained on $A$ and evaluated on $B$.
Figure~\ref{fig:original_vs_decoded_observables} reports representative results for dataset~$B$.

The top row compares the distributions of the maximum ADC value per event, resolved by detector layer and view, for the original data (left) and the decoded data after compression (right).
The bottom row shows the hit multiplicity per shower as a function of the particle momentum~$|p|$ for the same two cases.
Across all panels, the decoded distributions are indistinguishable from the originals within statistical precision, demonstrating exact preservation of both low-level detector readout and derived, physics-relevant observables.
These results confirm that the arithmetic coding procedure is fully lossless and invertible at the event level.
Importantly, the same level of agreement is obtained for both the unconditional codec and the codec conditioned on particle kinematics, indicating that invertibility is intrinsic to the coding scheme and independent of the specific factorization used in the probabilistic model.
This establishes arithmetic coding as a reliable compression--decompression mechanism suitable for downstream physics analysis, with no degradation of detector-level information.


\subsection{Compression Ratio Studies}
\label{subsec:compression_ratio}


To assess the practical performance of physics-aware arithmetic coding beyond its role as a fidelity diagnostic, we compare its lossless compression efficiency against a widely used general-purpose compressor, \texttt{gzip}. 
All comparisons are performed on identical raw content (see Eq. \eqref{eq:dataformat}), using a common canonicalized representation---\textit{i.e.}, a fixed, deterministic, and lossless serialization of the data---to ensure strict lossless and a fair, like-for-like evaluation across methods.
 Arithmetic coding is applied using the same physics-informed probabilistic factorization described earlier, while \texttt{gzip} is evaluated at multiple compression levels (\texttt{-1}, \texttt{-6}, and \texttt{-9}) \cite{gailly1992gnu}. The compression ratios reported in Table~\ref{tab:compression_AB_combined} are defined as the ratio of the original (uncompressed) data size to the compressed size, shown separately for the unconditional and kinematics-conditioned cases.
\begin{table}[!]
\centering
\caption{Lossless compression comparison.
Relative factors are quoted with respect to unconditional (U.-AC) and
conditional (C.-AC) arithmetic coding.}
\label{tab:compression_AB_combined}
\resizebox{\columnwidth}{!}{%
\begin{tabular}{lcccc}
\hline
Method & Size [MB] & Ratio & Rel.\ to U.-AC & Rel.\ to C.-AC \\
\hline
Uncompressed & 95.16 & -- & -- & -- \\
\textbf{U.-AC} & \textbf{7.02} & \textbf{13.55$\times$} & -- & 0.97$\times$ \\
\textbf{C.-AC} & \textbf{7.22} & \textbf{13.18$\times$} & 1.03$\times$ & -- \\
\texttt{gzip-9} & 11.21 & 8.49$\times$ & 1.60$\times$ & 1.55$\times$ \\
\texttt{gzip-6} & 11.86 & 8.02$\times$ & 1.69$\times$ & 1.64$\times$ \\
\texttt{gzip-1} & 15.25 & 6.24$\times$ & 2.17$\times$ & 2.11$\times$ \\
\hline
\end{tabular}}
\end{table}

Several conclusions emerge from these results.
Arithmetic coding consistently outperforms \texttt{gzip} at all compression levels. Even at the strongest \texttt{gzip} setting (\texttt{gzip-9}), arithmetic coding produces files that are approximately 1.6$\times$ smaller. At lower \texttt{gzip} levels, the advantage increases to nearly a factor of two. This demonstrates that generic compressors are unable to fully exploit the structured, physics-driven regularities present in detector data.
The achieved compression ratios indicate that arithmetic coding operates close to the Shannon limit implied by the chosen probabilistic representation. Because arithmetic coding maps probability assignments directly into code length, the resulting average description length provides a direct estimate of the cross-entropy between the data and the physics-informed factorized model. In contrast, \texttt{gzip} lacks an explicit probabilistic model and therefore cannot systematically approach this bound.
As shown in Table \ref{tab:entropy_unconditional} and \ref{tab:entropy_conditional}, while the calorimeter hit information dominates the compressed payload (approximately 90\% of the total size), particle kinematics still contribute non-negligibly (about 10\%). Importantly, including kinematic information incurs only a modest compression cost, while remaining essential for downstream fidelity, calibration, and consistency studies.

With respect to computational performance, we observe that the encoding time of the AC implementation is approximately a factor of two slower than that of \texttt{gzip}, while the decoding time is slower by a factor of $\mathcal{O}(10^2)$. 
This difference is expected: \texttt{gzip} relies on a highly optimized, production-grade implementation written in low-level compiled code, whereas the AC implementation used here is a reference, purely Python-based realization that prioritizes clarity, flexibility, and direct access to probabilistic structure over raw throughput. 
As such, the reported timing ratios should be interpreted as indicative rather than fundamental limitations of arithmetic coding itself; optimized implementations are well known to achieve substantially higher performance. 
A detailed optimization of encoding and decoding speed is therefore beyond the scope of the present work, whose focus is on compression efficiency and the information-theoretic interpretation of code length.

As we will expand (see Sec. \ref{sec:entropy_vs_codelength} and \ref{sec:budget}), our studies establish physics-aware arithmetic coding as more than an efficient lossless compressor. It provides a principled, interpretable, and near-optimal baseline for compression, while simultaneously enabling information-theoretic diagnostics of distributional fidelity. This dual role—as a compression method and as a quantitative scientific measurement tool—distinguishes arithmetic coding fundamentally from generic, task-agnostic compressors.

\subsection{Comparison between Shannon Entropy and Average Sequence Length}
\label{sec:entropy_vs_codelength}

Tables~\ref{tab:entropy_unconditional} and~\ref{tab:entropy_conditional} report the empirical entropy $H(q)$, the cross-entropy $H(p,q)$, and the achieved average code length $L$ (all in bits per event) for the unconditional and conditional codecs, respectively.
Results are shown separately for the calorimeter hits and for the particle-kinematics payload, as well as for the total event representation, and for both the training dataset $A$ and the validation $B$.
Several features are worth highlighting.

\begin{table}[!]
\centering
\caption{Shannon entropy, cross-entropy, and achieved average code length for the \emph{unconditional} arithmetic codec.}
\label{tab:entropy_unconditional}
\begin{tabular}{lcccc}
\toprule
Component & $H(q)$ & $H(p,q)$ & $L$ & Overhead (\%) \\
\midrule
(${A}$) Hits  & 755.01015 & 779.55374 & 779.55376 & $3.0\times10^{-6}$ \\
(${A}$) Part  &  82.41942 &  82.42861 &  82.42940 & $9.6\times10^{-4}$ \\
\midrule
(${A}$) Total & 837.42957 & 861.98234 & 861.98316 & $9.5\times10^{-5}$ \\
\midrule
(${B}$) Hits  & 749.72386 & 782.17784 & 782.17797 & $1.7\times10^{-5}$ \\
(${B}$) Part  &  82.41024 &  82.43129 &  82.43372 & $2.96\times10^{-3}$ \\
\midrule
(${B}$) Total & 832.13410 & 864.60912 & 864.61169 & $2.97\times10^{-4}$ \\
\bottomrule
\end{tabular}
\end{table}
\begin{table}[!]
\centering
\caption{Shannon entropy, cross-entropy, and achieved average code length for the \emph{conditional} arithmetic codec.}
\label{tab:entropy_conditional}
\begin{tabular}{lcccc}
\toprule
Component & $H(q)$ & $H(p,q)$ & $L$ & Overhead (\%) \\
\midrule
(${A}$) Hits  & 717.88968 & 798.01378 & 798.01380 & $2.0\times10^{-6}$ \\
(${A}$) Part  &  82.41942 &  82.42861 &  82.42940 & $9.6\times10^{-4}$ \\
\midrule
(${A}$) Total & 800.30910 & 880.44238 & 880.44320 & $9.3\times10^{-5}$ \\
\midrule
(${B}$) Hits  & 695.99810 & 806.13961 & 806.13969 & $1.1\times10^{-5}$ \\
(${B}$) Part  &  82.41024 &  82.43129 &  82.43372 & $2.96\times10^{-3}$ \\
\midrule
(${B}$) Total & 778.40833 & 888.57089 & 888.57342 & $2.8\times10^{-4}$ \\
\bottomrule
\end{tabular}
\end{table}

First, in all cases the achieved code length closely matches the empirical
cross-entropy, with discrepancies at the level of $\mathcal{O}(10^{-3})$ bits per event, corresponding to relative overheads below $10^{-3}\%$.
This demonstrates that the arithmetic codec operates essentially at the
Shannon-optimal limit for the specified probabilistic models.

Second, the cross-entropy evaluated on the validation dataset $B$ is systematically larger than that obtained on the training dataset $A$.
This increase is consistent with the expected additional ``surprise'' incurred
when encoding previously unseen data drawn from the same underlying process,
reflecting finite-sample estimation effects rather than a true distributional
difference, since the reference model is trained on $A$ rather than on $B$~\cite{mackay2003information}.

Third, conditioning the hit model on particle momentum reduces the empirical entropy $H(q)$ by capturing additional physical correlations.
At the same time, the cross-entropy $H(p,q)$ increases due to the added model structure and finite-sample effects within each conditioning bin.
Despite this increased complexity, the achieved code length remains tightly matched to $H(p,q)$, confirming that conditioning does not introduce any inefficiency in the coding procedure.

Overall, these results demonstrate that physics-aware arithmetic coding is not only lossless and invertible, but also achieves compression at the Shannon-optimal limit.
The excess code length $H(p,q)-H(p)=D_{\mathrm{KL}}(p\Vert q)$ provides a direct, information-theoretic measure of distributional mismatch, while the residual difference $L-H(p,q)$ quantifies the finite-precision overhead of the implementation.
This property underpins the use of arithmetic coding as a principled diagnostic for assessing distributional fidelity in high-dimensional scientific data.

\subsection{Information and Bit-Budget Decomposition}
\label{sec:budget}

Lossless compression provides a natural and physically interpretable decomposition of information content.
Because arithmetic coding assigns bits according to an explicit probabilistic factorization, the achieved codelength can be written as a sum of additive contributions associated with detector subsystems, readout channels, and modeled conditional structure.
This enables a fine-grained interpretation of where information resides in the data and which components dominate the overall description length.

\begin{figure}[!]
  \centering
  \includegraphics[width=\columnwidth]{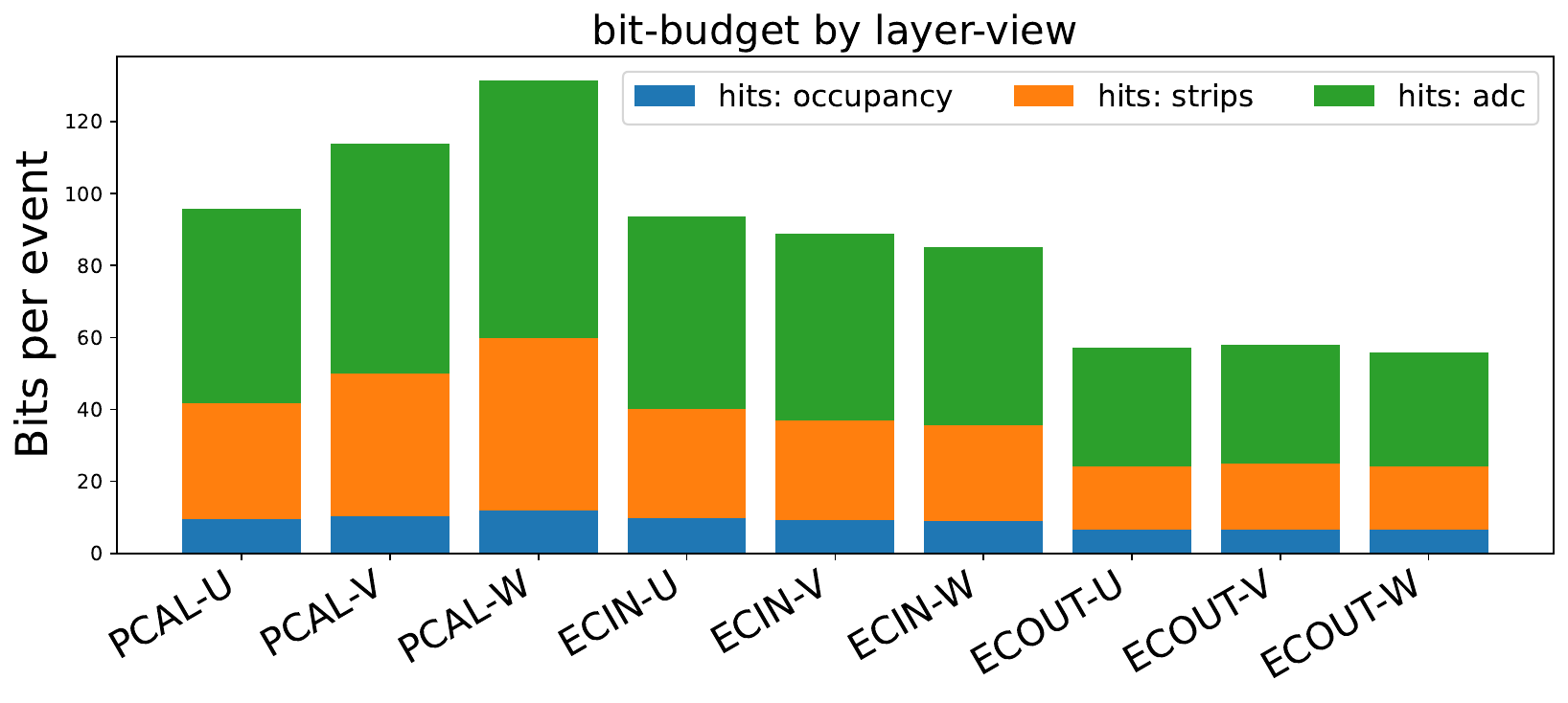}
  \caption{
    \textbf{Bit-budget decomposition for unconditional arithmetic coding.}
    The achieved codelength per event is decomposed by calorimeter layer and stereo view (U/V/W), and further split into occupancy, strip, and ADC contributions.
    The sum over all layer--views yields the total detector-level contribution to the achieved codelength.
  }
  \label{fig:bit_budget_A_uncond}
\end{figure}

Figure~\ref{fig:bit_budget_A_uncond} shows the achieved bit budget for the \emph{unconditional} arithmetic coding model trained on split~A, decomposed by calorimeter layer and stereo view.
For each layer--view (LV), the total contribution is further separated into occupancy, strip index, and ADC amplitude terms.
%
%
Table~\ref{tab:bit_budget_A_uncond} reports the corresponding numerical values.
The dominant contribution arises from ADC amplitudes, followed by strip indices, while occupancy carries a smaller but non-negligible fraction of the total information.
The PCAL layers contribute the largest share of the bit budget, reflecting both higher hit multiplicities and a broader ADC amplitude range, while ECIN and ECOUT contribute progressively less.
The final column reports the mean hit multiplicity per layer--view, illustrating the close correspondence between detector activity and information content.
Summing over all nine layer--views yields a detector-level contribution of approximately $779.55$~bits per event, which can be compared to Table~\ref{tab:entropy_unconditional}.
The particle kinematic information, encoded separately using a generic byte-level model, contributes an additional $82.43$~bits per event.
The resulting total achieved codelength of $\approx862$~bits per event defines the unconditional reference baseline.
Any possible reduction in codelength achieved by conditional or enriched probabilistic models can therefore be interpreted directly as recovered physical correlations, rather than changes in the coding algorithm.

\begin{figure}[!]
  \centering
  \includegraphics[width=\columnwidth]{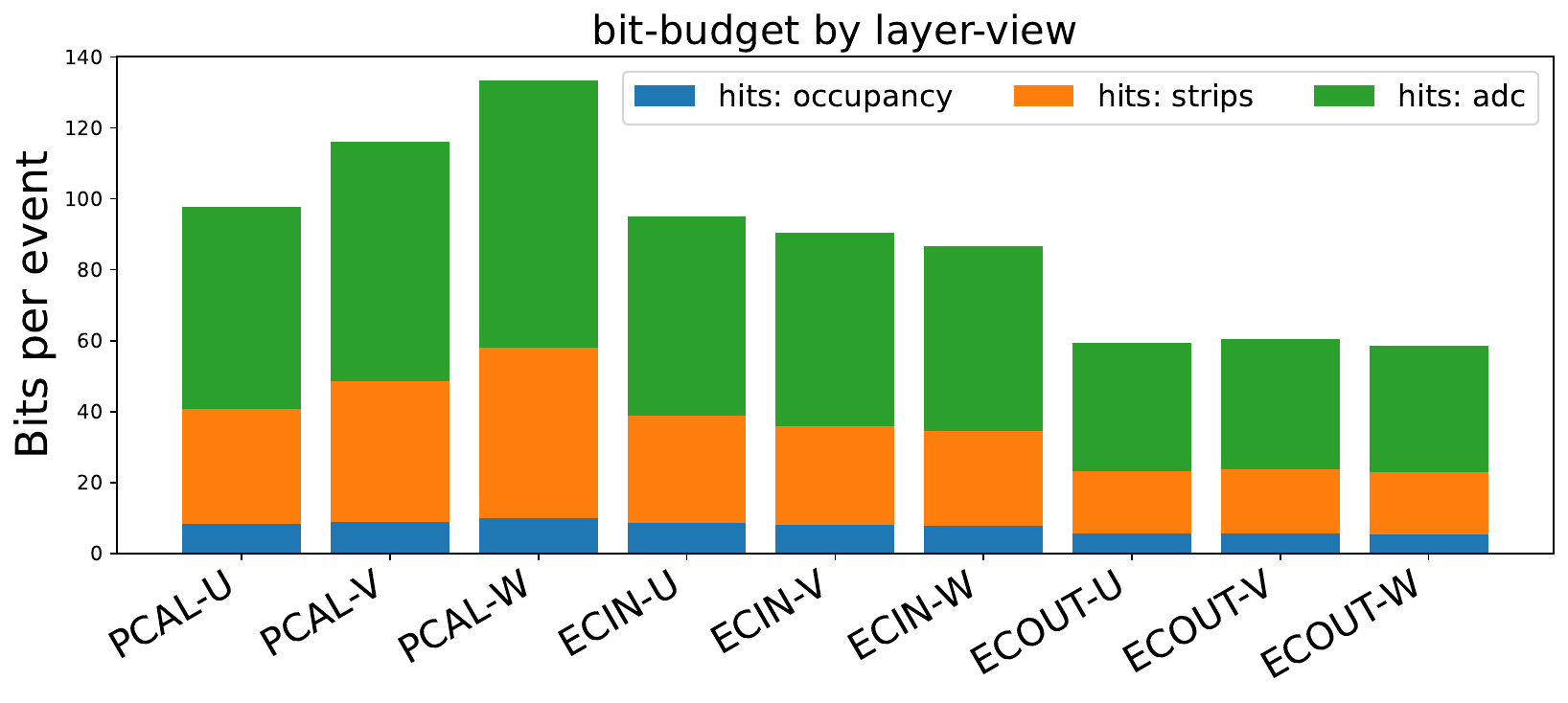}
  \caption{
    \textbf{Bit-budget decomposition for conditional arithmetic coding.}
    The achieved codelength per event is decomposed by calorimeter layer and stereo view (U/V/W), and further split into occupancy, strip, and ADC contributions.
    Conditioning the hit model on particle kinematics modifies the distribution of bits across layers while preserving the same additive decomposition.
  }
  \label{fig:bit_budget_A_cond}
\end{figure}

We now repeat the same information-theoretic decomposition for the \emph{conditional} arithmetic coding model, in which the detector hit probabilities are conditioned on bins of the particle momentum magnitude. 
Figure~\ref{fig:bit_budget_A_cond} and Table~\ref{tab:bit_budget_A_cond} show the achieved bit budget for split~A using this conditional codec.
Summing over all nine layer--views yields a detector-level contribution of approximately $798.01$~bits per event, which can be compared to Table~\ref{tab:entropy_conditional}.
Compared to the unconditional case, conditioning redistributes the bit budget across detector components.
In particular, the occupancy contribution is reduced across all layers, reflecting the fact that hit presence becomes more predictable once kinematic information is provided.
Conversely, the ADC contribution increases, indicating that amplitude distributions are evaluated within finer kinematic contexts, which incurs additional cross-entropy at finite statistics.
As a result, the total achieved codelength increases modestly.
This behavior is consistent with conditioning being applied only to the hit model, while particle kinematics are encoded separately using a fixed, generic representation.


\begin{table}[!]
\centering
\caption{
Per-layer--view decomposition of the achieved cross-entropy $H(q,p)$ for detector hits in the unconditional arithmetic coding model trained on split~A.
All values are reported in bits per event.
}
\label{tab:bit_budget_A_uncond}
\begin{tabular}{lrrrrr}
\toprule
LV & Occ. & Strip & ADC & Sum & $\langle N_{\text{hit}}\rangle$ \\
\midrule
PCAL--U  &  9.47 & 32.36 & 54.05 &  95.88 & 5.72 \\
PCAL--V  & 10.21 & 39.64 & 63.93 & 113.79 & 6.82 \\
PCAL--W  & 11.79 & 48.08 & 71.60 & 131.48 & 8.27 \\
ECIN--U  &  9.92 & 30.27 & 53.32 &  93.51 & 6.20 \\
ECIN--V  &  9.26 & 27.78 & 51.75 &  88.79 & 5.66 \\
ECIN--W  &  8.95 & 26.60 & 49.60 &  85.15 & 5.41 \\
ECOUT--U &  6.54 & 17.75 & 32.90 &  57.19 & 3.65 \\
ECOUT--V &  6.71 & 18.18 & 33.13 &  58.01 & 3.70 \\
ECOUT--W &  6.50 & 17.59 & 31.67 &  55.76 & 3.58 \\
\midrule
\multicolumn{4}{r}{Hits total} & 779.55 & \\
Particle kinematics & \multicolumn{3}{c}{---} & 82.43 & \\
\midrule
\textbf{Total} & \multicolumn{3}{c}{---} & \textbf{861.98} & \\
\bottomrule
\end{tabular}
\end{table}

\begin{table}[!]
\centering
\caption{
Per-layer--view decomposition of the achieved cross-entropy $H(q,p)$ for detector hits in the \emph{conditional} arithmetic coding model trained on split~A.
All values are reported in bits per event.
}
\label{tab:bit_budget_A_cond}
\begin{tabular}{lrrrrr}
\toprule
LV & Occ. & Strip & ADC & Sum & $\langle N_{\text{hit}}\rangle$ \\
\midrule
PCAL--U  &  8.44 & 32.36 & 57.03 &  97.83 & 5.72 \\
PCAL--V  &  8.84 & 39.64 & 67.75 & 116.23 & 6.82 \\
PCAL--W  &  9.90 & 48.08 & 75.45 & 133.43 & 8.27 \\
ECIN--U  &  8.63 & 30.27 & 56.03 &  94.93 & 6.20 \\
ECIN--V  &  8.08 & 27.77 & 54.61 &  90.46 & 5.66 \\
ECIN--W  &  7.81 & 26.59 & 52.39 &  86.80 & 5.41 \\
ECOUT--U &  5.50 & 17.75 & 36.09 &  59.34 & 3.65 \\
ECOUT--V &  5.62 & 18.17 & 36.76 &  60.55 & 3.70 \\
ECOUT--W &  5.42 & 17.59 & 35.45 &  58.46 & 3.58 \\
\midrule
\multicolumn{4}{r}{Hits total} & 798.01 & \\
Particle kinematics & \multicolumn{3}{c}{---} & 82.43 & \\
\midrule
\textbf{Total} & \multicolumn{3}{c}{---} & \textbf{880.44} & \\
\bottomrule
\end{tabular}
\end{table}

\subsection{Fidelity Studies}
\label{sec:fidelity}

We study the sensitivity of physics-aware lossless compression as a global diagnostic of distributional fidelity by constructing perturbed datasets with controlled strength.
Starting from an independent real-data split, we construct a family of calibration ``stress tests''
$C_{\varepsilon}$ by applying a controlled ADC scale perturbation
$a = 1+\varepsilon$ to occupied calorimeter hits, followed by rounding, clipping, and enforcement of the padding convention.
Each perturbed sample $C_{\varepsilon}$ is then compared to an independent baseline sample drawn from the same population.
All compression results use a \emph{fixed} reference codec trained once on the
training split (\textit{i.e.}, its CDFs are held fixed throughout the scan).
We report results for both the \emph{unconditional} codec and a \emph{conditional}
codec in which hit models are conditioned on $|p|$ via momentum bins, and we
compare against a standard kernel two-sample test based on the Maximum Mean
Discrepancy, evaluated on a fixed, physics-motivated feature representation.

\paragraph{Compression-based fidelity score under a fixed reference model.}

Let $q_A$ denote the probabilistic model (CDFs) learned from the reference sample~$A$ and then held fixed.
Following Eqs.~\eqref{eq:cross_entropy_empirical} and~\eqref{eq:deltaL_def}, for a dataset
$D=\{x_i\}_{i=1}^{N}$ the arithmetic codec induces an average per-event codelength
(in bits per event),
$L_A(D)$, up to negligible termination overhead
(Sec.~\ref{sec:entropy_vs_codelength}).
We define the \emph{excess codelength}
\begin{equation}
\Delta L(\varepsilon)
\;\equiv\;
L_A(C_{\varepsilon}) - L_A(B),
\label{eq:deltaL_def_pert}
\end{equation}
where $B$ is an independent baseline sample drawn from the same target distribution
as the unperturbed data used to construct $C_{\varepsilon}$.\footnote{
To minimize correlations between the perturbed sample and the reference used for hypothesis testing, we employ the three-way data-splitting strategy introduced in Sec.~\ref{sec:data}.
Briefly, the perturbation $C_{\varepsilon}$ is constructed from $B^{(1)}$,
the reference codec is trained once on $A^{(3)}$ (fixing $q_{A^{(3)}}$),
and excess codelengths are evaluated relative to an independent baseline sample $B^{(2)}$ drawn from the same target distribution.}
A positive $\Delta L$ indicates that events in $C_{\varepsilon}$ are, on average, less typical under the fixed reference model $q_A$ than events in the independent baseline sample $B$, \emph{i.e.}, they incur a larger cross-entropy under the same scoring rule.
This admits a likelihood-based interpretation: differences in mean codelength
correspond directly to differences in average negative log-likelihood evaluated
under the \emph{fixed} model $q_A$.
Accordingly, $\Delta L$ serves as a model-conditional log-likelihood \emph{score
difference} (expressed in bits per event), rather than a formal likelihood-ratio
test.

More generally, the same construction supports a distributional fidelity measure
between real and synthetic data.
Let $R$ denote a \emph{real} reference dataset used to define the probabilistic
representation of the codec, and let $S$ be a \emph{synthetic} dataset produced,
for example, by a generative model.
We define the \emph{fidelity gap}
\begin{equation}
\Delta L_{\mathrm{fid}}
\;\equiv\;
L_{R}(S)\;-\;L_{R}(R),
\label{eq:deltaL_fidelity}
\end{equation}
where both codelengths are evaluated under the same fixed reference model learned from $R$. 
A positive $\Delta L_{\mathrm{fid}}$ indicates that the synthetic sample is, on average, less typical than real data under the physical correlations encoded by the codec, while $\Delta L_{\mathrm{fid}}\simeq 0$ indicates that no statistically resolvable distributional mismatch is observed under the fixed probabilistic representation and available statistics.\footnote{In expectation, $\Delta L_{\mathrm{fid}}\ge 0$ when $R$ is drawn from the same distribution used to define the reference codec.
Negative values of $\Delta L_{\mathrm{fid}}$ may occur at finite sample size and
should be interpreted as statistical fluctuations.}

\paragraph{Blocked design and empirically calibrated test statistic.}
Because arithmetic coding produces a single compressed payload per dataset,
replicate measurements are obtained through a blocked design.
The baseline sample $B$ and the perturbed sample $C_{\varepsilon}$ are each
partitioned into $K$ disjoint blocks of equal size,
$\{B_k\}_{k=1}^K$ and $\{C_{\varepsilon,k}\}_{k=1}^K$.
For each block we compute the average codelength
\begin{equation}
L_A(D_k)\;\equiv\;
\frac{\,\lvert \mathrm{payload}_A(D_k)\rvert}{\lvert D_k\rvert}
\quad \text{(bits/event)},
\end{equation}
evaluated under the same fixed reference codec trained on $A$.
Let $\mu_C(\varepsilon)$ and $\mu_B$ denote the sample means of
$\{L_A(C_{\varepsilon,k})\}$ and $\{L_A(B_k)\}$, respectively,
with sample standard deviations $s_C(\varepsilon)$ and $s_B$.
We define the mean excess codelength
\begin{equation}
\Delta L(\varepsilon)
\;\equiv\;
\mu_C(\varepsilon) - \mu_B,
\end{equation}
and its associated standard error
\begin{equation}
\mathrm{SE}_{\Delta L}(\varepsilon)
\;=\;
\sqrt{\frac{s_C^2(\varepsilon)}{K} + \frac{s_B^2}{K}}.
\label{eq:deltaL_se}
\end{equation}
The corresponding standardized statistic is
\begin{equation}
t(\varepsilon)
\;=\;
\frac{\Delta L(\varepsilon)}{\mathrm{SE}_{\Delta L}(\varepsilon)}.
\label{eq:t_stat}
\end{equation}

Although Eq.~\eqref{eq:t_stat} coincides with a Welch-type test statistic under
Gaussian assumptions \cite{welch1947generalization}, we do not rely on its analytic reference distribution.
Instead, statistical significance is assessed using an \emph{empirical}
calibration of the null distribution of $t$, obtained from \emph{real}--vs--\emph{real} comparisons constructed under the same blocked procedure \cite{davison1997bootstrap, phipson2010}.
This ensures valid inference in the presence of finite-sample effects and
block-to-block heterogeneity.
Throughout this work we use $K=10$ blocks.

\paragraph{Comparison to MMD under a blocked design.}
As a complementary, model-agnostic baseline, we evaluate distributional fidelity
using the Maximum Mean Discrepancy (MMD) with a Gaussian RBF kernel~\cite{gretton2012kernel}.
To enable a statistically comparable analysis, we adopt the same blocked design
used for the compression-based tests.
Specifically, for each perturbation strength $\varepsilon$, we construct
$K$ approximately independent random blocks from the perturbed sample
$C_\varepsilon$ and from two independent real-data baselines $B^{(1)}$ and $B^{(2)}$
(Sec.~\ref{sec:data}).
For block index $k$, we define the blockwise contrast
\begin{equation}
d_k(\varepsilon)
=
\mathrm{MMD}^2\!\big(C^{(1)}_{\varepsilon,k},\,B^{(2)}_k\big)
-
\mathrm{MMD}^2\!\big(B^{(1)}_k,\,B^{(2)}_k\big),
\label{eq:mmd_block_diff}
\end{equation}
where $B^{(1)}_k$, $B^{(2)}_k$, and $C^{(1)}_{\varepsilon,k}$ are independently
constructed blocks of equal size.
The subtraction removes the intrinsic \emph{real}--vs--\emph{real} discrepancy arising from finite-sample fluctuations and isolates the excess MMD induced by the imposed perturbation, without requiring event- or block-level pairing.
The resulting blockwise differences $\{d_k(\varepsilon)\}_{k=1}^K$ are treated as approximately independent measurements.
Rather than relying solely on asymptotic analytic distributions, statistical
significance is assessed using an empirically calibrated null distribution
constructed from \emph{real}--vs--\emph{real} block comparisons.
This mirrors the empirical calibration procedure used for the compression-based
test and controls the false-positive rate under the null hypothesis of no
distributional difference.
Analytic one-sided $t$-statistics are used internally as test statistics to construct empirically calibrated $p$-values; the calibrated $p$-values are reported and used to assess sensitivity. Throughout this analysis we use $K=10$ blocks.

In practice, MMD is evaluated on a fixed-dimensional ($57$-dimensional) real-valued event representation.
For each of the nine calorimeter layer--view combinations, we compute (i) the occupancy fraction, defined as the fraction of the 20 hit slots with $\texttt{strip}\neq \texttt{pad\_strip}$, (ii) summary statistics of the ADC amplitudes over occupied slots (mean, standard deviation, and maximum), and (iii) summary statistics of the occupied strip indices (mean and standard deviation).
We then concatenate these $9\times 6=54$ features with the three components of the particle momentum, yielding 57 features per event.
This engineered representation balances physical interpretability and statistical conditioning of the kernel estimator.
As a result, MMD probes consistency only within this engineered fixed-dimensional summary space, whereas the compression-based test operates directly on the full discrete hit representation used by the codec.

\begin{figure}[t!]
    \centering
    \includegraphics[width=\linewidth]{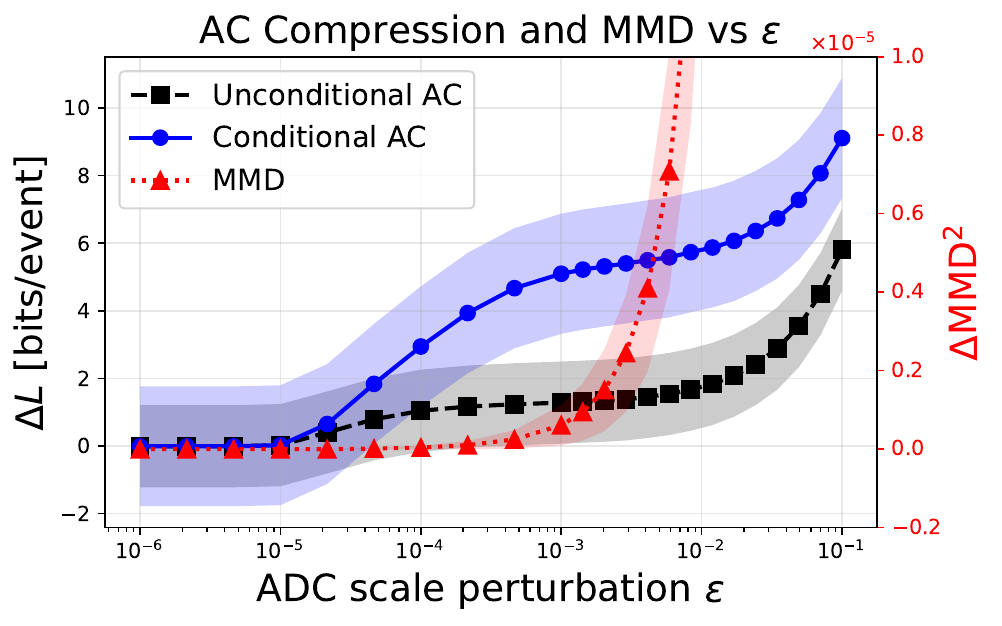}
    \caption{
    \textbf{Sensitivity of arithmetic coding and MMD to ADC scale perturbations~$\varepsilon$.}
    Mean excess codelength $\Delta L$ (left axis) for unconditional and conditional arithmetic coding is compared to the corresponding change in $\Delta \mathrm{MMD}^2$ (right axis).
    AC shows a smooth, monotonic response to increasing perturbations, while MMD remains relatively insensitive at small~$\varepsilon$ and increases sharply only at larger deviations.
    }   
    \label{fig:compression_vs_perturbation}
\end{figure}

\begin{figure}[t!]
    \centering
    \includegraphics[width=\linewidth]{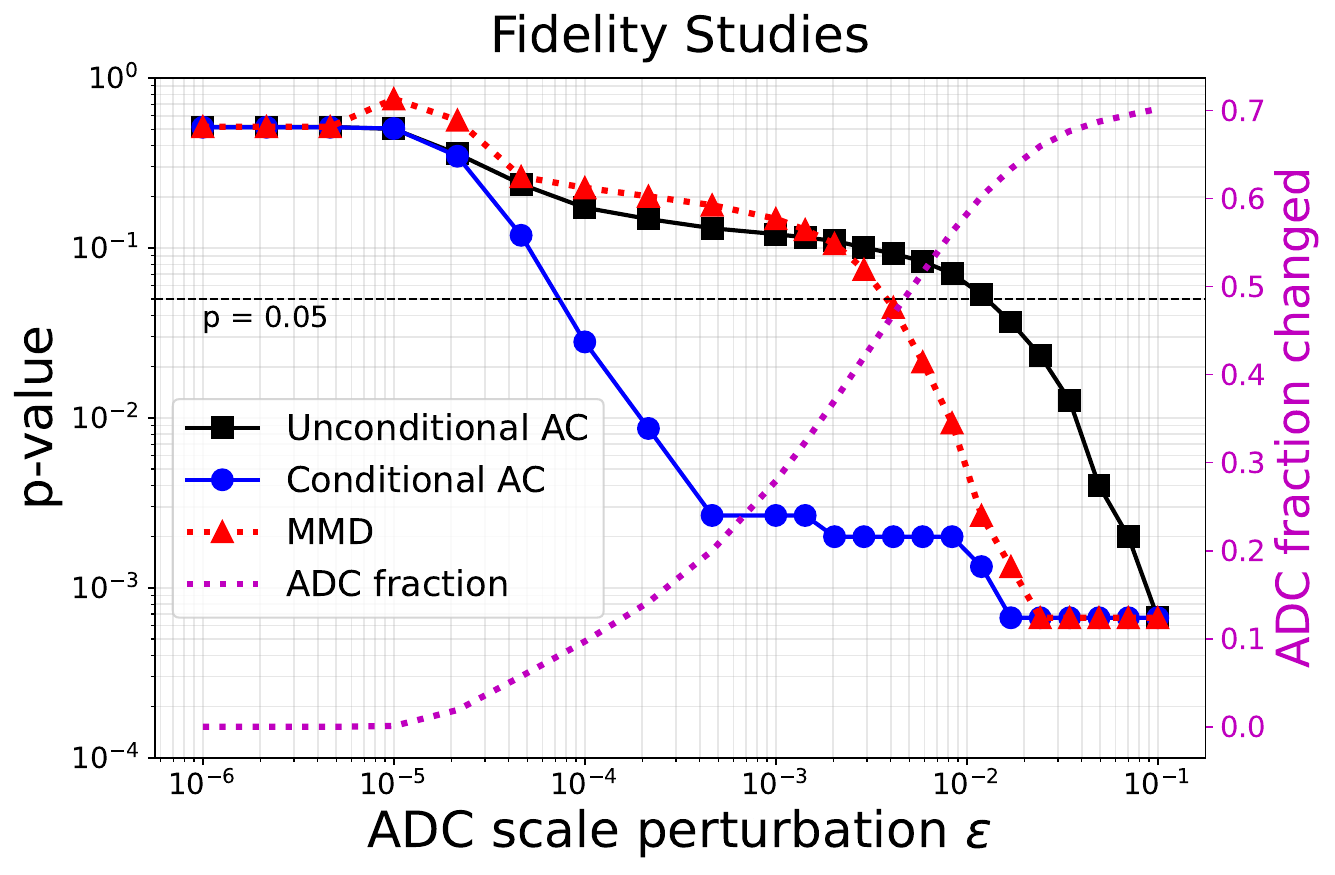}
    \caption{
    \textbf{Fidelity tests under controlled ADC scale perturbations~$\varepsilon$.}
    One-sided t-test $p$-values are shown for unconditional and conditional arithmetic coding (AC) and for the MMD-based test as a function of the ADC perturbation $\varepsilon$.
    The horizontal dashed line indicates the $p=0.05$ significance threshold.
    Conditional AC detects statistically significant deviations at substantially smaller $\varepsilon$ than MMD.
    The right axis reports the empirical fraction of ADC values that change under the perturbation, illustrating the relationship between physical modification rate and statistical sensitivity.  
    }
    \label{fig:pvalue_vs_perturbation}
\end{figure}

\paragraph{Results and interpretation.}
Figure~\ref{fig:compression_vs_perturbation} shows the mean excess codelength $\Delta L$ (left axis) for unconditional and conditional arithmetic coding, together with the change in $\Delta\mathrm{MMD}^2$ (right axis).
Both $\Delta L$ curves increase smoothly with $\varepsilon$, reflecting the monotonic increase in cross-entropy under the fixed reference codec as ADC symbols move across integer quantization boundaries and the heavy-tailed ADC distribution is progressively distorted.
Figure~\ref{fig:pvalue_vs_perturbation} reports the corresponding one-sided t-test $p$-values, with the conventional $p=0.05$ threshold indicated by the dashed line, and overlays the empirical fraction of ADC entries that change under the perturbation (right axis).
The conditional codec rejects the null hypothesis at $\varepsilon \simeq 1 \times 10^{-4}$, whereas the unconditional codec becomes significant only at substantially larger perturbations corresponding to $\varepsilon \gtrsim 10^{-2}$.
MMD exhibits a distinct sensitivity profile, remaining relatively flat at small $\varepsilon$ and then dropping sharply once the perturbation is large enough ($\varepsilon \gtrsim 4 \times 10^{-3}$) to induce a clear global shift in the kernel distance.
As mentioned, all fidelity tests are calibrated using an empirical null constructed from real data. As a result, the probability of falsely rejecting the null hypothesis when comparing \emph{real}--vs--\emph{real} is controlled at the nominal significance level $\alpha$. For $\alpha$ = 0.05, the conditional AC method becomes sensitive to perturbations starting at $\epsilon \sim \mathcal{O}(10^{-4})$.
Table~\ref{tab:big_comparison} summarizes the scan across $\varepsilon$, reporting $\Delta L$ and $p$-values for both codecs, as well as $\Delta\mathrm{MMD}^2$ and its corresponding $p$-value.

It is worth emphasizing that the compression-based and MMD-based diagnostics probe \emph{different null hypotheses}.
The compression test assesses whether $C_{\varepsilon}$ remains equally
\emph{typical} under a fixed, physics-informed reference model, as quantified by
its achieved codelength.
In contrast, MMD tests for equality of distributions in a reproducing kernel
Hilbert space, independent of any compression model.
As a result, a dataset may be statistically distinguishable from the reference
under MMD while remaining approximately compression-consistent with the fixed
physics-aware representation, and vice versa.

\begin{table*}[!]
\centering
\caption{Comparison of sensitivity metrics as a function of ADC scale perturbation $\varepsilon$.
For each method, we report the excess codelength $\Delta L$ for unconditional and conditional arithmetic coding,
the squared Maximum Mean Discrepancy $\Delta\mathrm{MMD}^2$, and the corresponding one-sided $p$-values.
Boldface indicates statistically significant deviations ($p<0.05$) for the corresponding method.}
\label{tab:big_comparison}
\begin{tabular}{ccccccc}
\toprule
$\varepsilon$
& $\Delta L$ (uncond)
& $p$ (uncond)
& $\Delta L$ (cond)
& $p$ (cond)
& $\Delta \mathrm{MMD}^2$
& $p$ (MMD) \\
\midrule
$1.00\times 10^{-6}$ & $-8.25\times 10^{-3}$ & $5.150\times 10^{-1}$ & $-1.11\times 10^{-2}$ & $5.130\times 10^{-1}$ & $0$ & $5.163\times 10^{-1}$ \\
$2.15\times 10^{-6}$ & $-8.25\times 10^{-3}$ & $5.150\times 10^{-1}$ & $-1.11\times 10^{-2}$ & $5.130\times 10^{-1}$ & $0$ & $5.163\times 10^{-1}$ \\
$4.64\times 10^{-6}$ & $-8.25\times 10^{-3}$ & $5.150\times 10^{-1}$ & $-1.11\times 10^{-2}$ & $5.130\times 10^{-1}$ & $0$ & $5.163\times 10^{-1}$ \\
$1.00\times 10^{-5}$ & $2.30\times 10^{-2}$ & $5.017\times 10^{-1}$ & $2.18\times 10^{-2}$ & $5.043\times 10^{-1}$ & $-1.69\times 10^{-10}$ & $7.535\times 10^{-1}$ \\
$2.15\times 10^{-5}$ & $3.98\times 10^{-1}$ & $3.578\times 10^{-1}$ & $6.49\times 10^{-1}$ & $3.464\times 10^{-1}$ & $-4.73\times 10^{-10}$ & $5.643\times 10^{-1}$ \\
$4.64\times 10^{-5}$ & $7.89\times 10^{-1}$ & $2.372\times 10^{-1}$ & $1.84$ & $1.186\times 10^{-1}$ & $1.37\times 10^{-8}$ & $2.625\times 10^{-1}$ \\
$1.00\times 10^{-4}$ & $1.04$ & $1.726\times 10^{-1}$ & $\mathbf{2.94}$ & $\mathbf{2.798\times 10^{-2}}$ & $4.02\times 10^{-8}$ & $2.258\times 10^{-1}$ \\
$2.15\times 10^{-4}$ & $1.17$ & $1.479\times 10^{-1}$ & $\mathbf{3.93}$ & $\mathbf{8.661\times 10^{-3}}$ & $1.02\times 10^{-7}$ & $2.012\times 10^{-1}$ \\
$4.64\times 10^{-4}$ & $1.23$ & $1.306\times 10^{-1}$ & $\mathbf{4.67}$ & $\mathbf{2.665\times 10^{-3}}$ & $2.48\times 10^{-7}$ & $1.785\times 10^{-1}$ \\
$1.00\times 10^{-3}$ & $1.28$ & $1.206\times 10^{-1}$ & $\mathbf{5.10}$ & $\mathbf{2.665\times 10^{-3}}$ & $6.11\times 10^{-7}$ & $1.486\times 10^{-1}$ \\
$1.43\times 10^{-3}$ & $1.31$ & $1.153\times 10^{-1}$ & $\mathbf{5.22}$ & $\mathbf{2.665\times 10^{-3}}$ & $9.49\times 10^{-7}$ & $1.279\times 10^{-1}$ \\
$2.03\times 10^{-3}$ & $1.34$ & $1.099\times 10^{-1}$ & $\mathbf{5.31}$ & $\mathbf{1.999\times 10^{-3}}$ & $1.50\times 10^{-6}$ & $1.059\times 10^{-1}$ \\
$2.89\times 10^{-3}$ & $1.38$ & $1.006\times 10^{-1}$ & $\mathbf{5.40}$ & $\mathbf{1.999\times 10^{-3}}$ & $2.45\times 10^{-6}$ & $7.462\times 10^{-2}$ \\
$4.12\times 10^{-3}$ & $1.45$ & $9.260\times 10^{-2}$ & $\mathbf{5.49}$ & $\mathbf{1.999\times 10^{-3}}$ & $\mathbf{4.10\times 10^{-6}}$ & $\mathbf{4.464\times 10^{-2}}$ \\
$5.88\times 10^{-3}$ & $1.54$ & $8.328\times 10^{-2}$ & $\mathbf{5.58}$ & $\mathbf{1.999\times 10^{-3}}$ & $\mathbf{7.08\times 10^{-6}}$ & $\mathbf{2.132\times 10^{-2}}$ \\
$8.38\times 10^{-3}$ & $1.67$ & $7.062\times 10^{-2}$ & $\mathbf{5.73}$ & $\mathbf{1.999\times 10^{-3}}$ & $\mathbf{1.26\times 10^{-5}}$ & $\mathbf{9.327\times 10^{-3}}$ \\
$1.19\times 10^{-2}$ & $1.83$ & $5.330\times 10^{-2}$ & $\mathbf{5.87}$ & $\mathbf{1.332\times 10^{-3}}$ & $\mathbf{2.30\times 10^{-5}}$ & $\mathbf{2.665\times 10^{-3}}$ \\
$1.70\times 10^{-2}$ & $\mathbf{2.08}$ & $\mathbf{3.664\times 10^{-2}}$ & $\mathbf{6.07}$ & $\mathbf{6.662\times 10^{-4}}$ & $\mathbf{4.31\times 10^{-5}}$ & $\mathbf{1.332\times 10^{-3}}$ \\
$2.42\times 10^{-2}$ & $\mathbf{2.42}$ & $\mathbf{2.332\times 10^{-2}}$ & $\mathbf{6.36}$ & $\mathbf{6.662\times 10^{-4}}$ & $\mathbf{8.20\times 10^{-5}}$ & $\mathbf{6.662\times 10^{-4}}$ \\
$3.46\times 10^{-2}$ & $\mathbf{2.88}$ & $\mathbf{1.266\times 10^{-2}}$ & $\mathbf{6.73}$ & $\mathbf{6.662\times 10^{-4}}$ & $\mathbf{1.58\times 10^{-4}}$ & $\mathbf{6.662\times 10^{-4}}$ \\
$4.92\times 10^{-2}$ & $\mathbf{3.55}$ & $\mathbf{3.997\times 10^{-3}}$ & $\mathbf{7.28}$ & $\mathbf{6.662\times 10^{-4}}$ & $\mathbf{3.09\times 10^{-4}}$ & $\mathbf{6.662\times 10^{-4}}$ \\
$7.02\times 10^{-2}$ & $\mathbf{4.49}$ & $\mathbf{1.999\times 10^{-3}}$ & $\mathbf{8.07}$ & $\mathbf{6.662\times 10^{-4}}$ & $\mathbf{6.05\times 10^{-4}}$ & $\mathbf{6.662\times 10^{-4}}$ \\
$1.00\times 10^{-1}$ & $\mathbf{5.80}$ & $\mathbf{6.662\times 10^{-4}}$ & $\mathbf{9.11}$ & $\mathbf{6.662\times 10^{-4}}$ & $\mathbf{1.19\times 10^{-3}}$ & $\mathbf{6.662\times 10^{-4}}$ \\
\bottomrule
\end{tabular}
\end{table*}

In this sense, compression-based fidelity is \emph{model-conditional}: enriching the probabilistic structure of the codec (\textit{e.g.},\ adding additional hit--kinematics correlations) changes the notion of fidelity being probed and can sharpen sensitivity.
As the sample size increases, MMD becomes asymptotically sensitive to arbitrarily small distributional differences, so its $p$-values quantify statistical detectability rather than practical or physical significance.
In contrast, because the compression-based test evaluates all samples under a
\emph{fixed} physics-informed reference model, the excess codelength
$\Delta L$ converges to a finite and interpretable measure of model mismatch,
expressed in bits per event.
Correlations between ADC values across strips are a fundamental feature of
electromagnetic showers, arising from lateral shower development, detector
geometry, and boundary effects in the readout.
Fidelity tests that operate only on marginal or per-channel distributions
therefore miss precisely the multi-channel structure that distinguishes
physically realistic detector data.
By operating directly on the joint detector readout, compression-based fidelity
tests remain sensitive to these correlations without requiring an explicit
event-level embedding.

Building on this sensitivity to joint detector structure, the compression-based
test evaluates whether a candidate dataset remains \emph{typical} under a fixed,
physics-informed probabilistic representation.
A large $p$-value in this framework does not constitute evidence that a dataset is ``real''; rather, it indicates that the dataset is statistically
indistinguishable from real data under the tested representation and available
statistics. Conversely, small $p$-values signal that the imposed perturbation induces systematic violations of the physical correlations encoded by the reference model.
The $\varepsilon$-scan therefore serves as a calibrated sensitivity study,
quantifying the smallest controlled distortions of the detector response that
can be reliably resolved while maintaining a controlled false-rejection rate on
unperturbed real data.
In this sense, the scan characterizes not only detectability, but also the
smallest perturbation scale that can be statistically resolved under the fixed
probabilistic representation and available statistics.

Overall, these studies demonstrate that physics-aware arithmetic coding provides
an interpretable, information-theoretic measure of distributional fidelity,
expressed directly in bits per event and equipped with a statistically principled decision threshold via blocked hypothesis testing.
Its behavior is complementary to kernel-based two-sample tests, which probe
distributional equality in an abstract feature space, whereas compression tests
consistency with respect to a fixed, physically motivated reference
representation.


\section{Conclusions}
\label{sec:conclusions}

In this work, we have demonstrated that physics-aware, lossless, and invertible
compression provides a principled and operational framework for both data reduction and quantitative fidelity assessment in multi-modal scientific datasets. We focused in particular on calorimeter detector readout, where structured correlations, discreteness, and detector symmetries play a central role. 
Using arithmetic coding with explicitly factorized probabilistic models,
we achieve compression performance that saturates the Shannon-optimal limit up
to negligible finite-precision overheads, while preserving exact invertibility
of the original detector-level information.

Beyond compression efficiency, a central result of this study is that achieved
codelengths admit a direct information-theoretic interpretation. Differences in
average codelength between datasets encoded under a fixed reference model
correspond to differences in cross-entropy, and therefore quantify
distributional mismatch in physically meaningful units of bits per event. In
this way, lossless compression becomes a global, model-conditional fidelity
diagnostic: datasets that remain statistically consistent with the same
underlying physics compress equally well, while deviations manifest as an
excess description length.

We have shown that even minimal probabilistic structure—encoding only essential
detector symmetries and factorization assumptions—already yields interpretable
``bit penalties'' that localize discrepancies to specific components of the
readout, such as calorimeter hits or kinematic payloads. This bit-budget
decomposition provides a transparent mapping between physical effects
(\textit{e.g.}, calibration distortions) and their information-theoretic cost,
enabling a level of interpretability that is often absent in black-box
statistical tests. At the event level, per-event codelengths define a meaningful
scalar quantity that can be aggregated for global fidelity assessment and, in
principle, leveraged to probe localized anomalies or rare mismodeling effects.

Through controlled perturbation studies, we demonstrated that
compression-based fidelity tests are sensitive to physically meaningful
detector distortions while remaining robust to transformations that preserve
the modeled structure. Compared to kernel-based distance measures such as MMD,
compression probes a complementary notion of fidelity: not whether two
distributions are identical in a generic feature space, but whether they remain
\emph{typical} under a fixed, physics-informed generative representation. This
distinction explains the differing sensitivity regimes observed and highlights
the value of combining information-theoretic and distributional diagnostics.
While conditional representations resolve additional physical correlations and
thereby reduce the intrinsic entropy of the data-generating process, their
achieved codelength can exceed that of unconditional models due to increased
modeling complexity and finite-sample effects. Nevertheless, our studies show
that conditional codecs exhibit enhanced sensitivity in perturbation scans,
indicating that richer structure amplifies statistically significant deviations
from the reference distribution.

The framework presented here naturally supports a train--test paradigm:
probabilistic models are learned on a reference dataset and then deployed
unchanged to evaluate independent samples, fast simulations, or systematically
perturbed data. In this setting, the excess codelength provides a single,
well-defined scalar observable with a clear statistical interpretation,
enabling hypothesis testing, calibration studies, and consistency checks
without reliance on hand-crafted observables or ad hoc distance measures.

Looking forward, several extensions are immediate. Richer conditional
structure—incorporating correlations with kinematics, timing information, or
event topology—can further sharpen sensitivity and refine the notion of
fidelity being tested. Event-level codelengths naturally support anomaly
detection and streaming applications, while integration with fast simulation
pipelines suggests a role for compression not only as a diagnostic, but as a
design and validation tool. More broadly, these results support the view that
lossless, physics-aware compression can serve as a foundational building block
for information-centric analysis of experimental data.


\begin{acknowledgments}
We thank Maurizio Ungaro (Jefferson Lab) and Richard Tyson (University of Glasgow) for generating and providing the Monte Carlo dataset used in this work. 
\end{acknowledgments}

\bibliographystyle{apsrev4-2}
\bibliography{references}

\end{document}